\documentstyle[12pt,epsf]{article}
\begin{document}
\def\thefootnote{\fnsymbol{footnote}}
\begin{flushright}
KANAZAWA-97-06\\ 
April, 1997
\end{flushright}
\vspace*{2cm}
\begin{center}
{\LARGE\bf $\mu \rightarrow e \gamma$ in supersymmetric multi $U(1)$
models with an abelian gaugino mixing}\\
\vspace{1 cm}
{\Large Daijiro Suematsu}
\footnote[1]{e-mail:suematsu@hep.s.kanazawa-u.ac.jp}
\vspace {1cm}\\
{\it Department of Physics, Kanazawa University,\\
        Kanazawa 920-11, Japan}\\    
\end{center}
\vspace{1cm}
{\Large\bf Abstract}\\  
A lepton flavor violating process $\mu\rightarrow e\gamma$ 
is investigated in the supersymmetric 
extra $U(1)$ models, which often appear as the low energy effective models
of superstring and can potentially solve the $\mu$-problem.
The branching ratio of this process is calculated.
It is numerically estimated and compared with that of the MSSM. 
In this study we
take account of an abelian gaugino kinetic term mixing and discuss
its influence on this process.
The possibility to find the extra gauge structure through
this process is discussed.
\newpage
\setcounter{footnote}{0}
\def\thefootnote{\arabic{footnote}}
The standard model(SM) has shown its incredible accuracy
to describe the electroweak interaction through the precise
measurement at LEP.
Nevertheless, physics beyond the SM is eagerly explored because
of its unsatisfactory feature for the explanation of the origin of 
weak scale and its stability.
Supersymmetrization of the SM is now considered
as the most promising extension to solve this problem\cite{n}.
However, even in this minimal supersymmetric standard model(MSSM) 
there still remains a theoretically unsatisfactory feature which
is known as the $\mu$-problem\cite{mu}. 
To cause an appropriate radiative symmetry breaking at the weak scale,
we need a Higgs mixing term $\mu H_1H_2$, where $\mu\sim O(G_F^{-1/2})$
and $G_F$ is a Fermi constant.
However, there is no reason why
$\mu$ should be such a scale because it is usually considered to be 
irrelevant to the supersymmetry breaking. 
A solution for this problem is to consider $\mu$ as a dynamical 
variable\cite{musol}.
The introduction of a singlet field $S$ with a Yukawa type coupling
$\lambda SH_1H_2$ can realize this scenario in the simplest way\cite{singlet}.
That is, if $S$ gets a vacuum expectation value(VEV) of order
1~TeV as a result of radiative 
corrections to the soft supersymmetry breaking parameters\cite{rad}, 
$\mu\sim O(G_F^{-1/2})$ will be realized dynamically 
through the relation $\mu=\lambda\langle S\rangle$.

The extra $U(1)$ models are the typical extensions of gauge 
structure of the SM.
It is very interesting to note that many extra $U(1)$ models 
have the above mentioned feature inevitably\cite{sy,extra}.
Low energy models derived from superstring often have accompanied with
extra $U(1)$ factors in their gauge structure\cite{string}.
It seems to be natural that these aspects motivate us 
to investigate extra $U(1)$
models and try to look for a clue of such a gauge structure.
Recent precise measurements at LEP and also the Tevatron experiments show 
us that the lower bound for its gauge boson mass is rather
large and then it may not be so easy to find it directly\cite{exp}.
Even in that case if nature is supersymmetric and the
gauge bosons have their superpartners, there may be other
possibilities to investigate the
gauge structure through examining the processes to which their
superpartners contribute.

In this letter we study the lepton flavor violating 
$\mu \rightarrow e\gamma$ process.
The gauginos of extra $U(1)$s can affect this process.
Our purpose here is to estimate their effect and discuss the possibility 
to find the extra gauge structure through this process.  
Its comparison with the results in the MSSM will also be useful for the 
future experimental analysis.
We consider the minimal models which have only one extra $U(1)_X$ and
a singlet Higgs $S$ with a $U(1)_X$ charge besides the MSSM
contents.\footnote{In order to induce the symmetry breaking
radiatively, it is necessary to introduce the vector like extra color
triplets $(g, \bar g)$ which have the coupling to the singlet $S$ as
$\kappa Sg \bar g$\cite{sy}. But in the present study 
they play no role and then we
will ignore them.}  
These fields are assumed to remain light around the TeV region.
The neutralino sector in this model is extended by an extra
$U(1)_X$ gaugino and a fermionic partner of the singlet Higgs $S$ in
addition to the ingredients of the MSSM.

Before proceeding to the detailed study we should note an additional
feature of the neutralino sector of these extra $U(1)$ models.
It has been well known that in principle there can be kinetic term mixings
among abelian gauge fields because these field strengthes are gauge
invariant.\footnote{In fact there are some works in which it is discussed 
in what case kinetic term mixings can occur\cite{mixing,mixing2}.}
Supersymmetrization of the models introduces kinetic term mixings
among abelian gauginos. In the analysis of multi $U(1)$ models
we generally need to take account of these effects.
Thus at first we briefly summarize the mixing effects in the gaugino sector
for the usage in the later study.

In supersymmetric models gauge fields are extended to vector superfields
\begin{equation}
V_{\rm WZ}(x, \theta, \bar\theta)=-\theta\sigma_\mu\bar\theta V^\mu
+i\theta\theta\bar\theta\bar\lambda-i\bar\theta\bar\theta\theta\lambda
+{1\over 2}\theta\theta\bar\theta\bar\theta D,
\end{equation}
where we used the Wess-Zumino gauge.
A gauge field strength is included in the chiral superfield
constructed from $V_{\rm WZ}$ in the well known procedure,
\begin{eqnarray}
W_\alpha(x, \theta)&=&(\bar D\bar D)D_\alpha V_{\rm WZ} \nonumber \\
&=&4i\lambda_\alpha-4\theta_\alpha
D+4i\theta^\beta\sigma_{\nu\alpha\dot\beta}
\sigma^{\dot\beta}_{\mu\beta}(\partial^\mu V^\nu-\partial^\nu V^\mu)
-4\theta\theta\sigma_{\mu\alpha\dot\beta}\partial^\mu\bar\lambda^{\dot\beta}.
\end{eqnarray}
In terms of these superfields the supersymmetric Lagrangian can be written as
\begin{equation}
{\cal L}={1\over 32}\left(W^\alpha W_\alpha\right)_F
+\left(\Phi^\dagger \exp(2g^0QV_{\rm WZ})\Phi\right)_D, 
\end{equation}
where $\Phi=(\phi, \psi, F)$ is the chiral superfield representing 
matter fields.
This Lagrangian is easily extended to multi $U(1)$ models.
In the models with two $U(1)$ factor groups,
the supersymmetric gauge invariant kinetic terms are most generally
written by using chiral 
superfields $\hat W_\alpha^a$ and $\hat W^b_\alpha$ 
for $U(1)_a\times U(1)_b$ as\cite{mixing2},
\begin{equation}
{1\over 32}\left(\hat W^{a\alpha} \hat W^a_\alpha\right)_F
+{1\over 32}\left(\hat W^{b\alpha} \hat W^b_\alpha\right)_F
+{\sin\chi\over 16}\left(\hat W^{a\alpha} \hat W_{\alpha}^b\right)_F
\end{equation}
where we introduced the mixing terms.
These can be canonically diagonalized by performing the transformation,
\begin{equation}
\left(\begin{array}{c} \hat W^a \\ \hat W^b \\ \end{array}\right)
=\left(\begin{array}{cc}1 & -\tan\chi \\ 0 & 1/\cos\chi \\ 
\end{array}\right)\left(\begin{array}{c} W^a \\ W^b \\
\end{array}\right).
\end{equation}
This transformation affects not only the gauge vector
fields but also the sector of gauginos $\lambda_{a,b}$ and 
auxiliary fields $D_{a,b}$. 
The modification due to this transformation 
in the gaugino sector can be summarized as
\begin{equation}
g_a^0Q_a\hat \lambda^a+g_b^0Q_b\hat
\lambda^b=g_aQ_a\lambda^a+\left(g_{ab}Q_a+g_bQ_b\right)\lambda^b,
\end{equation}
where $\lambda_{a,b}$ are canonically normalized gauginos.
$Q_a$ and $Q_b$ stand for the charges of matter fields for $U(1)_a$
and $U(1)_b$. Gauge coupling constants $g_a, g_{ab}$ and $g_b$ are
related to the original ones as,
\begin{equation}
g_a=g_a^0, \quad g_{ab}=g_a^0\tan\chi, \quad g_b={g_b^0\over\cos\chi}.
\end{equation}
These low energy values are determined by using the renormalization
group equations. However, in the present study we will treat them as 
parameters.

For the study of the $\mu\rightarrow e\gamma$ process in the
supersymmetric models, it is necessary to clarify both of the
neutralino and chargino sector.
The relevant part of the superpotential and soft supersymmetry breaking
terms are
\begin{eqnarray}
W&=&\lambda SH_1H_2 +\cdots, \nonumber\\
{\cal L}_{\rm soft}&=&-\sum_im_i^2|\phi_i|^2 \nonumber\\
&+&{1\over2}(M_W\lambda_W\lambda_W
+M_Y\lambda_Y\lambda_Y+M_X\lambda_X\lambda_X
+M_{YX}\lambda_Y\lambda_X +{\rm h.c.})+\cdots,
\end{eqnarray}
where $\phi_i$ represents the scalar components contained in the models.
$M_W,~M_Y$ and $M_X$ are soft supersymmetry breaking gaugino
masses\footnote{
We introduce the effect caused from the 
abelian gaugino mass mixing as $M_{YX}$, which may exist at the 
Planck and may also be yielded through the loop effects.
We need to estimate its low energy value by using the
renormalization group equations. In the later numerical study we put 
$M_{YX}=0$, for simplicity.} 
for $SU(2)_L$, $U(1)_Y$ and $U(1)_X$.
These parameters and a Yukawa coupling $\lambda$ are assumed to be real and it 
should not be confused with the gaugino fields $\lambda_a$.
Using the canonically normalized basis, we can write down the relevant 
quantities in the neutralino sector modified by the kinetic term
mixing. They are the neutralino mass matrix and the vertex factors 
of gaugino-fermion-sfermion interactions.
If we take the canonically normalized gaugino basis as 
${\cal N}^T=(-i\lambda_{W_3}, -i\lambda_Y, 
-i\lambda_X, \tilde H_1, \tilde H_2, \tilde S)$ and define
the mass terms as 
$$
{\cal L}_{\rm mass}^n=-{1\over 2}{\cal N}^T{\cal MN}+{\rm h.c.},
$$
the 6 $\times$ 6 neutralino mass matrix ${\cal M}$ can be expressed as
\begin{equation}
\left( \begin{array}{cccccc}
M_W & 0 & 0 &m_Zc_W\cos\beta & -m_Zc_W\sin\beta &0 \\
0 & M_Y & C_1 & -m_Zs_W\cos\beta & m_Zs_W\sin\beta &0 \\
0 & C_1 & C_2 & C_3 & C_4 & C_5 \\
m_Zc_W\cos\beta &-m_Zs_W\cos\beta &  C_3& 0 & 
\lambda u & \lambda v\sin\beta \\
-m_Zc_W\sin\beta & m_Zs_W\sin\beta &C_4 & \lambda u &
0 & \lambda v\cos\beta \\
0 & 0& C_5 & \lambda v\sin\beta & \lambda v\cos\beta & 0\\
\end{array} \right).
\end{equation}
Matrix elements $C_1\sim C_5$ are components which are affected by the 
kinetic term mixing. They are represented as
\begin{eqnarray}
&&C_1=-M_Y\tan\chi +{M_{YX}\over\cos\chi},\quad
C_2=M_Y\tan^2\chi+{M_X \over
\cos^2\chi}-{2M_{YX}\sin\chi\over\cos^2\chi},  \nonumber \\
&&C_3={1\over \sqrt 2}\left(g_Y\tan\chi+{g_XQ_1\over\cos\chi}\right)
v\cos\beta,\quad
C_4={1\over \sqrt 2}\left(-g_Y\tan\chi+{g_XQ_2\over\cos\chi}\right)
v\sin\beta,\nonumber\\
&&C_5={1\over \sqrt 2}{g_XQ_S\over\cos\chi}u,
\end{eqnarray}
where $Q_1, Q_2$ and $Q_S$ are the extra $U(1)_X$ charges of Higgs
chiral superfields $H_1, H_2$ and $S$.

Neutralino mass eigenstates $\tilde\chi_i^0(i=1\sim 6)$ are related 
to ${\cal N}_j$
through the mixing matrix $U$ as
\begin{equation}
\tilde\chi_i^0=\sum_{j=1}^6 U_{ij}^T{\cal N}_j.
\end{equation}
The change induced by the kinetic term mixing 
in the gaugino interactions can be confined into 
the extra $U(1)_X$ gaugino sector and by using eq.(6) 
new interaction terms can be expressed as,
\begin{eqnarray}
&&{i \over \sqrt 2}\left[\tilde \psi^*\left(-g_YY\tan\chi+{g_XQ_X\over
\cos\chi}\right)\lambda_X\psi-\left(-g_YY\tan\chi+{g_XQ_X\over\cos\chi}
\right)\bar\lambda_X\bar \psi\tilde \psi\right. \nonumber \\
&&\left.+H^*\left(-g_YY\tan\chi+{g_XQ_X\over
\cos\chi}\right)\lambda_X\tilde
H-\left(-g_YY\tan\chi+{g_XQ_X\over\cos\chi}
\right)\bar\lambda_X\bar{\tilde H}H\right]
\end{eqnarray} 
where $\psi$ and $\tilde \psi$ represent the quarks/leptons and 
the squarks/sleptons, respectively.
Higgs fields $ (H_1, H_2, S)$ are summarized as $H$ and the corresponding
Higgsinos $\tilde H_1$, $\tilde H_2$ and $\tilde S$ are denoted as 
$\tilde H$. 
Taking account of this, gaugino-fermion-sfermion vertecies 
in the basis of mass eigenstates 
are assigned by the following factors,
\begin{eqnarray}
&&Z_i^L(Y, Q_X)=-{1\over \sqrt 2}\left[g_W\tau_3U_{1i}+g_YYU_{2i}+
\left(-g_YY\tan\chi+{g_XQ_X\over \cos\chi}\right)U_{3i}\right], \nonumber \\
&&\overline{Z_i^R}(Y, Q_X)={1\over \sqrt 2}\left[g_YYU_{2i}+
\left(-g_YY\tan\chi+{g_XQ_X\over \cos\chi}\right)U_{3i}\right],
\end{eqnarray}
where we used the left handed basis for the chiral superfields.
It is also necessary to define the chargino mass eigenstates for the
following calculation.
The chargino mass term is given in the matrix form as
\begin{equation}
{\cal L}_{\rm mass}^c=-\left(H_1^-, -i\lambda^-_Y\right)
\left(\begin{array}{cc}-\lambda u& \sqrt 2m_Zc_W\cos\beta\\
\sqrt 2m_Zc_W\sin\beta& M_W \\ \end{array}\right)
\left(\begin{array}{c}H_2^+ \\ -i\lambda^+_Y\\ \end{array}\right) +{\rm h.c.}.
\end{equation}
The mass eigenstates are defined in terms of the weak interaction 
eigenstates through unitary transformations,
\begin{equation}
\left(\begin{array}{c}\tilde \chi_1^+\\ \tilde \chi_2^+\\ \end{array}\right)
=W^{(+)\dagger}\left(\begin{array}{c}H_2^+\\ -i\lambda^+_Y\\
\end{array}\right),  \qquad
\left(\begin{array}{c}\tilde \chi_1^-\\ \tilde \chi_2^-\\ \end{array}\right)
=W^{(-)\dagger}\left(\begin{array}{c}H_1^-\\ -i\lambda^-_Y\\ 
\end{array}\right).
\end{equation}

Based on these preparations,  we proceed to the estimation 
of the $\mu\rightarrow e\gamma$
process in the present models.
The flavor changing processes are
strongly suppressed through the experimental results\cite{fcncexp}. 
In the supersymmetric models, however, there are generally 
many sources for these 
processes in the superpartner sector besides the ones of the 
SM[13-16].
Colored superpartners cause the dominant contributions
in many hadronic flavor changing neutral processes.
Thus the contribution from the neutralino sector may not be clearly 
seen through such processes.
In order to see the structure of the neutralino sector,
we need non-hadronic process and  $\mu\rightarrow e\gamma$ seems to
be particurally interesting in the relation to our present purpose
as far as R-parity violating terms are absent.
At one-loop level this process can occur because of the existence of
nontrivial flavor structure of soft supersymmetry breaking terms in
the slepton sector.
Various studies of $\mu\rightarrow e\gamma$ in the MSSM framework 
and some extended models have been done by now\cite{susyfcnc,leptonv,lepfcnc}.
We extend these analyses to the multi $U(1)$s case.
One-loop diagrams contributing to this process in the present models 
are shown in Fig.1.

The effective interaction describing this decay is given as 
\begin{equation}
{\cal L}_{\rm eff}={\cal G}_L\bar\psi_{e_R}\sigma_{\mu\nu}\psi_{\mu_L} 
F^{\mu\nu}+{\cal G}_R\bar\psi_{e_L}\sigma_{\mu\nu}\psi_{\mu_R} 
F^{\mu\nu}. 
\end{equation}
By carrying out the calculation of diagrams in Fig.1, 
we can obtain the effective couplings ${\cal G}_L$ and ${\cal G}_R$. 
In the diagonalizing basis of the lepton mass matrix $m_l$,
the origin of flavor 
changings is the off-diagonal elements of Kobayashi-Maskawa matrix
in the lepton sector\footnote{We assume the non-zero Majorana neutrino 
masses induced from the seesaw mechanism in view of the solar
neutrino problem.}
and also the slepton mass matrix.
The slepton mass matrix is written as\footnote{The origin of flavor violating
off-diagonal elements of $M_{LL}^2, M_{RR}^2$ and $M_{LR}^2$ are discussed 
from various view points\cite{leptonv,lepfcnc}.
In this analysis we donot refer to it and only treat them as
parameters.}
\begin{equation}
\left(\begin{array}{cc}
M_{LL}^2& M_{LR}^2\\ M_{RL}^2& M_{RR}^2
\end{array}\right)
\end{equation}
where $M_{LR}^2=m_l(A + \lambda u\tan\beta)$ for charged sleptons.
In the neutrino sector right-handed sneutrinos are assumed to have the 
large supersymmetric masses and their relevant part to the present analysis 
is only the part of $M_{LL}^2$.
To reduce the number of free parameters
we make the following assumptions for components of the slepton mass 
matrix,
\begin{equation}
(M_{LL}^{e,\nu})^2_{\alpha\alpha}=(M_{RR}^{e})^2_{\alpha\alpha}
\equiv M^2, \qquad
(M_{LL}^{e,\nu})^2_{\alpha\beta}=(M_{RR}^{e})^2_{\alpha\beta}
\equiv \Delta_{\alpha\beta}^2.
\end{equation}
As shown in Fig.1, there are two types of diagrams which are distinguished by
the place of the chirality flip. 
For our present purpose, it will be enough to concentrate our
attention on 
the diagrams with the chirality flip on the internal line ( Figs
(a) and (c)).\footnote{
This treatment may not be bad even in the quantitative view point
since the neutralino masses are expected to be much larger than
charged lepton masses.
For the completeness of our formulus, however, 
we will present the contribution to the effective couplings
from Figs. (b) and (d) in the appendix.}
Under these assumptions the effective couplings 
${\cal G}_L$ and ${\cal G}_R$ can be summarized as
\begin{eqnarray}
&&\hspace{-7mm}{\cal G}_L=-{e\over 32\pi^2}
\left[~\sum_{i=1}^6 m_i^n\left\{
{(M_{LR}^e)^2_{\mu e}\over M^4}\left(Z_{2i}^L(-1,Q_{e_L})
\overline{Z_i^R}(2,Q_{e_R})+
{g_W^2m_\mu m_e\over 2m_W^2\cos^2\beta}U_{4i}^2 \right)
\right.\right.\nonumber\\
&&\left. -{g_W \over \sqrt 2m_W\cos\beta}
{\Delta^2_{\mu e}\over M^4}
\left(m_e Z_{2i}^L(-1,Q_{e_L})
+m_\mu \overline{Z_{i}^R}(2,Q_{e_R})\right)U_{4i}\right\}
F_1({m_i^{n2}\over M^2}) \nonumber\\
&&\left.-{g_W^2m_e\over \sqrt 2m_W\cos\beta}
\sum_{i=1}^2\left(
K_{\nu_e\mu}{m_i^c \over M^2} J({m_i^{c2}\over M^2})+\sum_{\alpha=\mu,\tau}
K_{\nu_\alpha\mu}{\Delta^2_{\alpha e}\over m_i^{c3}}
F_1({M^2\over m_i^{c2}})\right)
W_{2i}^{(+)}W_{1i}^{(-)}~\right],\nonumber \\
&&\hspace{-7mm}{\cal G}_R=-{e\over 32\pi^2}
\left[~\sum_{i=1}^6 m_i^n\left\{
{(M_{RL}^e)^2_{\mu e}\over M^4}\left(Z_{2i}^L(-1,Q_{e_L})
\overline{Z_i^R}(2,Q_{e_R})+
{g_W^2m_\mu m_e\over 2m_W^2\cos^2\beta}U_{4i}^2 \right)
\right.\right.\nonumber\\
&&\left. -{g_W \over \sqrt 2m_W\cos\beta}
{\Delta^2_{\mu e}\over M^4}
\left(m_\mu Z_{2i}^L(-1,Q_{e_L})
+m_e \overline{Z_{i}^R}(2,Q_{e_R})\right)U_{4i}\right\}
F_1({m_i^{n2}\over M^2}) \nonumber\\
&&\left.-{g_W^2m_\mu\over \sqrt 2m_W\cos\beta}
\sum_{i=1}^2\left(K_{\nu_\mu e}{m_i^c \over M^2} J({m_i^{c2}\over M^2})+
\sum_{\alpha=e,\tau}
K_{\nu_\alpha e}{\Delta^2_{\alpha\mu}\over m_i^{c3}}
F_1({M^2\over m_i^{c2}})\right)
W_{2i}^{(+)}W_{1i}^{(-)}~\right], 
\end{eqnarray}
 where $m_i^n$ and $m_i^c$ represent the i-th mass eigenvalues of
neutralinos and charginos, respectively.
$\Delta^2_{\mu e}$ stands for an off-diagonal element between 
the $e$- and $\mu$-generation of slepton mass
matrices as defined by eq.(18).
Its allowed range may be estimated at a few GeV$^2$ or less
depending on other soft supersymmetry breaking parameters\cite{fcnc}. 
$K_{\alpha\beta}$ is the Kobayashi-Maskawa matrix element 
in the lepton sector. 
Kinematical functions $F_1(r)$ and $J(r)$ appearing 
from the loop integrals are defined by
\begin{eqnarray}
&&F_1(r)={1\over 2(1-r)^4}\left[~ 1+4r-5r^2+2r(r+2)\ln r ~\right],\nonumber\\
&&J(r)={1\over 2(1-r)^2}\left[-3+r-{2\over (1-r)}\ln r\right].
\end{eqnarray}
Using these results, we can represent the branching ratio of this
decay process as
\begin{equation}
B(\mu\rightarrow e\gamma)={48\pi^2\over G_F^2 m_\mu^2}
\left(~\vert{\cal G}_L\vert^2+\vert{\cal G}_R\vert^2\right).
\end{equation}

In order to compare this result with the MSSM one,
it is useful to list up the extra parameters added to the ones 
contained in the MSSM formulus:
$$\tan\beta,~ 
(M_{LR}^{e})^2_{\alpha\beta},~ M^2,~\Delta^2_{\alpha\beta},~
 K_{\alpha\beta},~ M_W,~ M_Y.$$
Additional parameters to these are new gaugino masses $(M_X, ~M_{YX})$, 
the kinetic term mixing parameter $\sin\chi$, 
the extra $U(1)$ coupling $g_X$ and
charges\footnote{It should be noted that $Q_1+Q_2+Q_S=0$ is satisfied
because of the form of superpotential.}
$(Q_1, Q_2)$ and also the $\mu$-term relevant parameters 
$(\lambda, \langle S\rangle(\equiv u))$.
We can easily check that the neutralino contribution to eq.(21)
results in the expression given in refs.\cite{susyfcnc,fcnc}
in the case of the photino dominated neutralino, 
if we put these additional parameters zero instead of keeping
$\mu(=\lambda u)$ constant.

Before choosing a parameter set for the numerical analysis, 
we should note some features of our models.
In these models the vacuum expectation value $u$ of the singlet Higgs $S$
is relevant to the extra $U(1)_X$ gauge boson mass besides determining the
$\mu$-scale. The mixing between the ordinary $Z^0$ and the $U(1)_X$
boson is severely constrained by the precise measurement at
LEP and the direct search at Tevatron\cite{exp}.
This constraint requires that the mass of the $U(1)_X$ boson is large
enough\footnote{The mixing component of the mass matrix can be small
enough for the special value of $\tan\beta$ and $\sin\chi$.
In such a case this requirement is not necessary to be satisfied.}
and in that case its mass eigenvalue is given
by\cite{sy}
\begin{equation}
m_{Z^\prime}^2 \simeq {1\over 2\cos^2\chi}g_X^2(Q_1^2v_1^2+Q_2^2v_2^2+
Q_S^2u^2).
\end{equation}
The experimental bound on $m_{Z^\prime}^2$ determines the lower bound
on $u$. 
On the other hand, $u$ determines the $\mu$-scale as $\mu=\lambda u$. 
Thus to keep $\mu$ in the suitable range we need to put the upper
bound on $\lambda$.
For its rough estimation, we take $m_{Z^\prime}~{^>_\sim}~400$ GeV and
also assume $g_X=g_Y$, which is satisfied, for example, in the abelian 
subgroup of $E_6$ if the full components of ${\bf 27}$ of $E_6$
contribute to $\beta$-functions. In this case if we require
$\mu~{^<_\sim}~1$ TeV, we obtain the upper bound on $\lambda$ as
$\lambda~{^<_\sim}~0.6Q_S$.
This bound seems to be reasonable from the view point of the analysis
of the radiative symmetry breaking\cite{sy}.

Another feature which we should note is the dependence of the
effective couplings ${\cal G}_L$ and  ${\cal G}_R$ on
the neutralino and chargino mass eigenvalues $m_i^n$ and $m_i^c$.
These dependence can be factorized as $r^{1/2}F_1(r)$ where
$r=(m_i^n/M)^2$ for neutralinos and also $r^{-3/2}F_1(r^{-1})$, 
$r^{1/2}J(r)$ where $r=(m_i^c/M)^2$ for charginos. 
They vary in the range ~$0< r^{1/2}F_1(r)~{^<_\sim}~0.1$,~ 
$0< r^{-3/2}F_1(r^{-1})~{^<_\sim}~0.25$ and $0<r^{1/2}J(r)~{^<_\sim}~0.44$. 
Each maximum value is realized at
$r\sim 0.27$, $r\sim 0.025$ and $r\sim 0.12$, respectively.
Thus all of these factors can be considered as the same order 
at least except for $r\sim 0$.

Taking account of this, we can roughly estimate the condition for the
neutralino contribution dominance by comparing the neutralino
contribution to the branching ratio with the chargino contribution.
The couplings of the neutralinos to leptons come from gauge
couplings and Yukawa couplings.
Because Yukawa couplings are small enough, the dominant contribution
will be yielded by the neutralinos which are dominantly composed of
the gauginos $\lambda_W, \lambda_Y$ and $\lambda_X$.
As seen from eq.(19), it is naively required 
\begin{equation}
{\Delta^2_{\mu e}\over (M_{LR}^e)_{\mu e}^2}~{^>_\sim}~
{m_W\cos\beta \over m_\mu g_W }\sim 10^3
\end{equation}
in order to guarantee the similar order contribution from all neutralinos.
This requirement seems to be difficult to be satisfied.
So we focus our study to the case where main neutralino
contribution comes from
the term associated with $(M_{LR}^e)_{\mu e}^2$ in eq.(19).
Next we compare it with the chargino contribution.
If we pay attention on the factors in each terms of eq.(19)
except for the mixing matrix elements,
the condision for the neutralino contribution becoming larger than the 
chargino one can be roughly estimated as
\begin{equation}
(M_{LR}^e)^2_{\mu e}~{^>_\sim}~{m_\mu\over m_W\cos\beta}
K_{\nu_\mu e}M^2 \sim 20K_{\nu_\mu e}.
\end{equation}
In this estimation we assumed $\tan\beta\sim 1$ and $M\sim 100$~GeV.
If we note that $(M_{LR}^{e})^2_{\mu e}=m_\mu A_{\mu e}$ and 
 $m_l\mu\tan\beta$ does not contribute to it, the above condition
for $(M_{LR}^{e})^2_{\mu e}$ corresponds to 
$A_{\mu e}~{^>_\sim}~200K_{\nu_\mu e}~{\rm GeV}$.
Thus if we take $K_{\nu_\mu e}\sim 5\times 10^{-4}$ as the KM-matrix
element in the lepton sector,\footnote{This small value does not
contradict the neutrino oscillation solution for the solar neutrino
problem, if we assume the existence of a sterile neutrino as
ref.\cite{neut}.} 
the neutralino 
contribution assumed above can be expected to be dominant
under the condition,\footnote{These values are very similar to ones
given in ref.\cite{fcnc}.}
\begin{equation}
 \Delta_{\mu e}<1~{\rm GeV},\qquad A_{\mu e}~{^>_\sim}~10^{-1}~{\rm
 GeV}.
\end{equation}
This argument suggests that the gaugino components of the neutralino
contribution can be dominant one in the
rather general situation.
Moreover, under this condition $B(\mu\rightarrow e\gamma)$ takes 
the value just below the present experimental bound.
Thus in such a parameter range we may have a chance
to get a hint for an additional abelian gauge structure 
through the $\mu\rightarrow e \gamma$ process.  
Our main interest is the effect coming from new ingredients so that 
in the following numerical study we assume the measure of flavor mixing 
$(M_{LR}^{e})^2_{\mu e}$ so as $B(\mu\rightarrow e\gamma)$ to be 
the same order as the present experimental bound in the MSSM case.

Now we give our result of the numerical analysis.
As the typical values of free parameters, we take
\begin{eqnarray}
&&\tan\beta=1.5, \qquad A_{\mu e}=0.2~{\rm GeV}, 
\qquad M=100~{\rm GeV}, \nonumber\\
&&M_Y=M_X={5\over 3}\tan^2\theta_WM_W,\qquad M_{YX}=0, \qquad \lambda=0.5,
\end{eqnarray} 
where we assumed the unification relation for the gaugino masses.
As a typical example of the low energy extra $U(1)$,
we take the $\eta$-model induced from $E_6$.
Their charge 
assignment for the relevant fields is listed in Table 1.
\begin{figure}[t]
\begin{center}
\begin{tabular}{|c||c|c|c|c|c|c|c|c|}\hline
fields & $Q$ & $U^c$ & $D^c$ & $L$ & $E^c$ & $H_1$ & $H_2$ &$S$ \\\hline\hline
$Y$ & ${1\over 3}$  & $-{4\over 3} $ &${2\over 3}$& $-1$ &2 &$-1$ &1 
& 0\\\hline
$Q_{\eta}$& $-{2\over3}$&$-{2\over3}$&${1\over3}$&${1\over3}$&$-{2\over3}$&
${1\over3}$&${4\over3}$&$-{5\over3}$\\\hline
\end{tabular}
\vspace*{.3cm}\\
{\small {\bf Table 1} \hspace{.3cm}
The charge assignments of extra $U(1)$s induced from $E_6$.
These charges are normalized as $\displaystyle \sum_{i \in {\bf 27}}Q_i=20$.
Only relevant fields to our study are listed from ${\bf 27}$ of $E_6$.}
\end{center}
\end{figure}
Under this parameter setting, in Fig.2 
$B(\mu\rightarrow e\gamma)$ in this model is drawn as a function of $u$ 
in the case of $M_W=80,~180$~ GeV and $\sin\chi=0,~0.3$. 
The horizontal axis
should be converted to $u$ by $u=50(u^\prime+2)$.
Thus the range of $u$ is $100~{\rm GeV} \le u\le 2000~{\rm GeV}$.
Here it is useful to recall again that $m_{Z^\prime}$ is given by eq.(22) and
also the relation $\mu=\lambda u$.
From the recent chargino and neutralino search at LEP\cite{aleph},
the allowed region of $(\mu,~M_W)$ plane should satisfy 
$\mu,~M_W~{^>_\sim}~100$~GeV as far as $\mu >0$.
This corresponds to $u^\prime >2$ for $\lambda=0.5$.
Moreover, taking account of the small mixing contraint on $Z^0$ and
the extra $U(1)$, 
$u \ge 800$~GeV and then $u^\prime \ge 14$ should be satisfied 
if we require $m_{Z^\prime} \ge 400$~GeV,.

The ratio $R$ of this $B(\mu\rightarrow e\gamma)$ 
against the one of the MSSM is also presented in Fig.3 
in the case of $\sin\chi=0$ and 0.3 for each value of $M_W$.
As easily seen from Figs.2 and 3, the influence appearing in 
$B(\mu\rightarrow e\gamma)$ due to the extra 
$U(1)$ gaugino is not so large but non-negligible. 
These figures also show that the gaugino kinetic term mixing 
has an enhancement effect on $B(\mu\rightarrow e\gamma)$ in the 
$\eta$-model.
This effect is considered to be mainly caused from the change of the effective 
couplings $Z_i^L(Y,Q_X)$ and $\overline{Z_i^R}(Y,Q_X)$.
The smaller $M_W$ and $u$ give the larger $B(\mu\rightarrow e\gamma)$.
The effect of $\sin\chi\not=0$ can be easily seen for the smaller 
$M_W$ and $u$.

It should be noted that there can remain the deviation of $O(10^{-12})$
 from the MSSM value at $u^\prime \sim 14$.
As $u^\prime$ becomes larger, $B(\mu\rightarrow e\gamma)$
monotonically decreases but the $O(10^{-13})$ deviation can still
remain.
For the smaller $u(u^\prime <14)$, the suitable condition should 
be satisfied for the consistency with the bound of the mixing 
between $Z^0$ and the extra $U(1)_X$, as already remarked in the footnote.
if such a condition is satisfied and then the small value of $u$ is
allowed from the precise measurement at LEP, we may have an
important clue to study the extra gauge structure because of the
large deviation from the MSSM prediction.
Anyway, these results seem to be interesting.
Particularly, the fact that $B(\mu\rightarrow e\gamma)$ can 
deviate by $O(10^{-12\sim -13})$ from the MSSM value even
at the rather large $u$ region seems to be encouraging.
It may be possible to find some clues of the extra $U(1)$ gauge structure 
through $B(\mu\rightarrow e\gamma)$ if its experimental bound is improved by
an order from the present value.

Some comments should be ordered on the various parameter dependences of
$B(\mu\rightarrow e\gamma)$. In particular, $\lambda$ and $\tan\beta$
dependence seems to be important.\\
1.~In the present models $\lambda$ can be an independent parameter
and it affects $B(\mu\rightarrow e\gamma)$ through
the neutralino mass matrix directly besides through the
combination with $u$.
As a result, $B(\mu\rightarrow e\gamma)$ shows the 
substantial $\lambda$ dependence, although no significant $\lambda$
dependence can be seen in the ratio $R$ at the large $u$ region.\\ 
2.~The $\tan\beta$ dependence seems to be very crucial for the
absolute value of $B(\mu\rightarrow e\gamma)$ in the whole range of
$u$. The large $\tan\beta$ makes $B(\mu\rightarrow e\gamma)$ small for 
all $u$ region and also increases the sensitivity to $u$ at the 
small $u$ region.
No significant $\tan\beta$ dependence can be seen in $R$ at the large
$u$ region.\\
We may also be interested in the $M_{YX}$ dependence in the case of
$\sin\chi\not=0$. As far as $M_{YX}$ is induced by the loop effects, 
$M_{YX}$ may be roughly estimated
as $M_{YX}\sim g_Yg_{YX}M_Y/8\pi^2$, which is completely 
negligible. Unless $M_{YX}$ is produced at $M_{\rm pl}$ as a rather
large value, it can be safely neglected in this type of study.

In summary we investigated the $\mu\rightarrow e\gamma$ process in the 
extra $U(1)$ models taking account of the gaugino kinetic term
mixing. After driving the formulus for the branching ratio
of $\mu\rightarrow e\gamma$
in the general framework, we practiced the numerical study and 
showed that in the $\eta$-model the deviation 
from the MSSM can be seen at the level of $O(10^{-12\sim -13})$
through this process. 
The abelian gaugino kinetic term 
mixing has some effects on this process and we may find the suitable
clue of the extra gauge structure by investigating this process.
It will be useful to note that the usage of this process 
may open the alternative window to 
search the extra $U(1)$ gauge structure for an appropriate parameter
region.
It will be necessary to investigate this process in more general
parameter region and other extra $U(1)$ models. 
The combined study with other rare process related to the neutralino 
sector like the electron electric dipole moment is also interesting\cite{edme}.

\vspace{.5cm}

This work is partially supported by a Grant-in-Aid for Scientific
Research from the Ministry of Education, Science and Culture
(\#08640362).

\newpage

{\Large\bf Appendix}
\vspace{.5cm}

The contribution to the effective couplings ${\cal G}_L$ and 
${\cal G}_R$ coming from Figs.(b) and (d) are summarized as,
\begin{eqnarray}
&&\hspace{-7mm}{\cal G}_L=-{em_\mu\over 32\pi^2}
\left[\sum_{i=1}^6 F_3({m_i^{n2}\over M^2})
\left\{{(M_{LR}^e)^2_{\mu e}\over M^4}
{g_W(m_\mu+m_e) \over \sqrt 2m_W\cos\beta}Z_{2i}^L(-1,Q_{e_L})U_{4i}
\right.\right.\nonumber\\
&&\hspace{30mm}\left.+{\Delta^2_{\mu e}\over M^4}
\left(Z_{2i}^L(-1,Q_{e_L})Z_{2i}^L(-1,Q_{e_L})+
{g_W^2m_\mu m_e\over 2m_W^2\cos^2\beta}U_{4i}^2\right)\right\}
 \nonumber\\
&&\hspace{30mm}+\sum_{i=1}^2 
g_W^2F_4({m_i^{c2}\over
M^2})\left(\sum_{\alpha,\beta(\alpha\not=\beta)}^{e,\mu,\tau}
K_{\nu_\alpha \mu}
{\Delta^2_{\alpha\beta}\over M^4}K_{\nu_\beta e}^\dagger\right)
W_{2i}^{(+)\ast}W_{2i}^{(+)}\Bigg], \nonumber \\
&&\hspace{-7mm}{\cal G}_R=-{em_\mu\over 32\pi^2}
\left[\sum_{i=1}^6 F_3({m_i^{n2}\over M^2})
\left\{ {(M_{LR}^e)^2_{\mu e}\over M^4}
{g_W(m_\mu+m_e) \over \sqrt 2m_W\cos\beta}
\overline{Z_i^R}(2,Q_{e_R})U_{4i}\right.\right.\nonumber\\
&&\hspace{30mm}\left.+{\Delta^2_{\mu e}\over M^4}
\left(\overline{Z_i^R}(2,Q_{e_R})\overline{Z_i^R}(2,Q_{e_R})+
{g_W^2m_\mu m_e\over 2m_W^2\cos^2\beta}U_{4i}^2\right)\right\}
 \nonumber\\
&&\hspace{30mm}+\sum_{i=1}^2 
g_W^2F_4({m_i^{c2}\over M^2}){\Delta^2_{\mu e}\over M^4}
{m_\mu m_e\over 2m_W^2\cos^2\beta} 
W_{1i}^{(-)\ast}W_{1i}^{(-)}\Bigg], 
\end{eqnarray}
where we used the fact $m_\mu \gg m_e$ and the kinematical functions
$F_3(r)$ and $F_4(r)$ are defined by
\begin{eqnarray}
&&F_3(r)={1\over 12(1-r)^5}\left[-1+9r+9r^2-17r^3+6r^2(r+3)\ln
r\right],\nonumber \\
&&F_4(r)={1\over 6(1-r)^5}\left[1+9r-9r^2-r^3+6r(r+1)\ln r\right].
\end{eqnarray}
In the limit of $m_i^n\rightarrow 0$, $B(\mu\rightarrow e\gamma)$
calculated from the photino contribution in ${\cal G}_L$ can
be easily checked to be reduced to the result given in ref.\cite{susyfcnc}.


\newpage
\noindent
{\Large\bf Figure Captions}
\vspace{1cm}\\
{\Large\bf Fig. 1}
\vspace{.3cm}\\
One-loop diagram contributing to the effective coupling ${\cal G}_L$
of $\mu_L\rightarrow e_R\gamma$.
Figs.(a) and (b) represent the neutralino contribution and Figs.(c)
and (d) represent the chargino contribution. 
There are similar diagrams in which the chirality of the external
lines are exchanged.
Flavor mixings are induced by the
off-diagonal elements of slepton mass matrices which are expressed
by $\bullet$.  It should be noted that the chirality flip occurs on
the internal line in (a) and (c) and on the external line
in (b) and (d).
\vspace{1cm}\\
{\Large\bf Fig. 2}
\vspace{.3cm}\\
$B(\mu\rightarrow e\gamma)$ as a function of $u$ in the $\eta$-model.
The vertical axis $Br$ stands for $10^{11}\times B(\mu\rightarrow
e\gamma)$ and the horizontal axis $u^\prime$ should be understood as
$u=50(u^\prime+2)$. 
Each line corresponds to the various parameter settings for $(M_W,~
\sin\chi)$ and their values are taken as $A(80,~ 0)$, $B(180,~ 0)$, 
$C(80,~ 0.3)$ and $D(180, ~0.3)$.
\vspace{1cm}\\
{\Large\bf Fig. 3}
\vspace{.3cm}\\
The ratio $R$ of $B(\mu\rightarrow e\gamma)$ against the MSSM  as a
function of $u$ in $\eta$-model.
$R$ is defined as $R=B_\eta(M_W, \sin\chi)/B_{\rm MSSM}(M_W)$.
Each line corresponds to the various parameter settings for $(M_W,~
\sin\chi)$ and their values are taken as $A(80,~ 0)$, $B(180,~ 0)$, 
$C(80,~ 0.3)$ and $D(180, ~0.3)$.
\newpage
\pagestyle{empty}
\vspace*{2cm}
\setlength{\unitlength}{1mm}
\begin{center}
\begin{picture}(72,50)(0,0)
\thicklines
\put(35,3){\bf (a)}
\put(63,44){${\cal A}_\mu$}
\put(12,44){$\mu_L$}
\put(63,10){$e_R^c$}
\put(23,36){${\cal V}_{\mu_L}$}
\put(50,11){${\cal V}_{e_R^c}$}
\put(29,44){($\tilde \mu_L, \tilde\mu_R^c$)}
\put(56,26){($\tilde e_R^c, \tilde e_L$)}
\put(41,18){$\tilde\chi^0$}
\put(29,30){$\tilde\chi^0$}
\put(38,27){$\bigoplus$}
\put(42,40){\circle*{3}}
\put(8,40){\vector(1,0){10}}
\put(18,40){\line(1,0){10}}
\multiput(28,40)(5,0){5}{\line(1,0){3}}
\put(28,40){\line(1,-1){5}}
\put(40,28){\vector(-1,1){7}}
\put(40,28){\vector(1,-1){7}}
\put(52,16){\line(-1,1){5}}
\multiput(52,16)(0,5){5}{\line(0,1){3}}
\multiput(53,40)(4,0){5}{\oval(2,3)[t]}
\multiput(55,40)(4,0){5}{\oval(2,3)[b]}
\put(72,16){\vector(-1,0){10}}
\put(52,16){\line(1,0){10}}
\end{picture}\hspace{1cm}
\begin{picture}(72,50)(0,0)
\thicklines
\put(35,3){\bf (b)}
\put(63,44){${\cal A}_\mu$}
\put(12,44){$\mu_L$}
\put(63,10){$e_R^c$}
\put(25,34){${\cal V}_{\mu_R}$}
\put(50,11){${\cal V}_{e_R^c}$}
\put(28,44){$(\tilde \mu_R^c, \tilde\mu_L)$}
\put(20,44){$\mu_R^c$}
\put(56,26){$(\tilde e_R^c, \tilde e_L)$}
\put(34,23){$\tilde\chi^0$}
\put(43,40){\circle*{3}}
\put(8,40){\vector(1,0){5}}
\put(13,40){\line(1,0){10}}
\put(28,40){\vector(-1,0){5}}
\put(16,39){$\bigotimes$}
\multiput(28,40)(5,0){5}{\line(1,0){3}}
\put(28,40){\vector(1,-1){14}}
\put(52,16){\line(-1,1){10}}
\multiput(52,16)(0,5){5}{\line(0,1){3}}
\multiput(53,40)(4,0){5}{\oval(2,3)[t]}
\multiput(55,40)(4,0){5}{\oval(2,3)[b]}
\put(72,16){\vector(-1,0){10}}
\put(52,16){\line(1,0){10}}
\end{picture}\vspace{1cm}\\
\begin{picture}(72,50)(0,0)
\thicklines
\put(35,3){\bf (c)}
\put(63,10){${\cal A}_\mu$}
\put(15,44){$\mu_L$}
\put(65,44){$e_R^c$}
\put(32,43){$\tilde \nu_{\mu_L}$}
\put(46,43){$\tilde \nu_{e_L}^\ast$}
\put(56,26){$\tilde\chi^-$}
\put(34,23){$\tilde\chi^+$}
\put(25,34){${\cal V}_{\mu_L}$}
\put(54,35){${\cal V}_{e_R^c}$}
\put(8,40){\vector(1,0){10}}
\put(18,40){\line(1,0){10}}
\multiput(28,40)(5,0){5}{\line(1,0){3}}
\put(52,16){\vector(-1,1){16}}
\put(28,40){\line(1,-1){10}}
\put(44,21){$\bigoplus$}
\put(52,40){\line(0,-1){10}}
\put(52,16){\vector(0,1){14}}
\put(52,40){\line(1,0){10}}
\put(72,40){\vector(-1,0){10}}
\put(42,40){\circle*{3}}
\multiput(53,16)(4,0){5}{\oval(2,3)[t]}
\multiput(55,16)(4,0){5}{\oval(2,3)[b]}
\end{picture}\hspace{1cm}
\begin{picture}(72,50)(0,0)
\thicklines
\put(35,3){\bf (d)}
\put(63,10){${\cal A}_\mu$}
\put(10,44){$\mu_L$}
\put(63,44){$e_R^c$}
\put(32,44){$\tilde\nu_{\mu_L}$}
\put(22,44){$\mu_R^c$}
\put(46,43){$\tilde \nu_{e_L}^\ast$}
\put(56,26){$\tilde\chi^-$}
\put(34,23){$\tilde\chi^-$}
\put(25,34){${\cal V}_{\mu_R}$}
\put(54,34){${\cal V}_{e_R^c}$}
\put(8,40){\vector(1,0){5}}
\put(13,40){\line(1,0){10}}
\put(28,40){\vector(-1,0){5}}
\put(16,39){$\bigotimes$}
\multiput(28,40)(5,0){5}{\line(1,0){3}}
\put(52,16){\line(-1,1){14}}
\put(28,40){\vector(1,-1){12}}
\put(52,40){\line(0,-1){10}}
\put(52,16){\vector(0,1){14}}
\put(62,40){\line(-1,0){10}}
\put(72,40){\vector(-1,0){10}}
\put(42,40){\circle*{3}}
\multiput(53,16)(4,0){5}{\oval(2,3)[t]}
\multiput(55,16)(4,0){5}{\oval(2,3)[b]}
\end{picture}
\vspace*{3cm}\\
{\Large\bf Fig. 1}
\end{center}

\newpage
\begin{figure}[htbp]
\begin{center}
\epsfxsize=\textwidth
\epsfbox{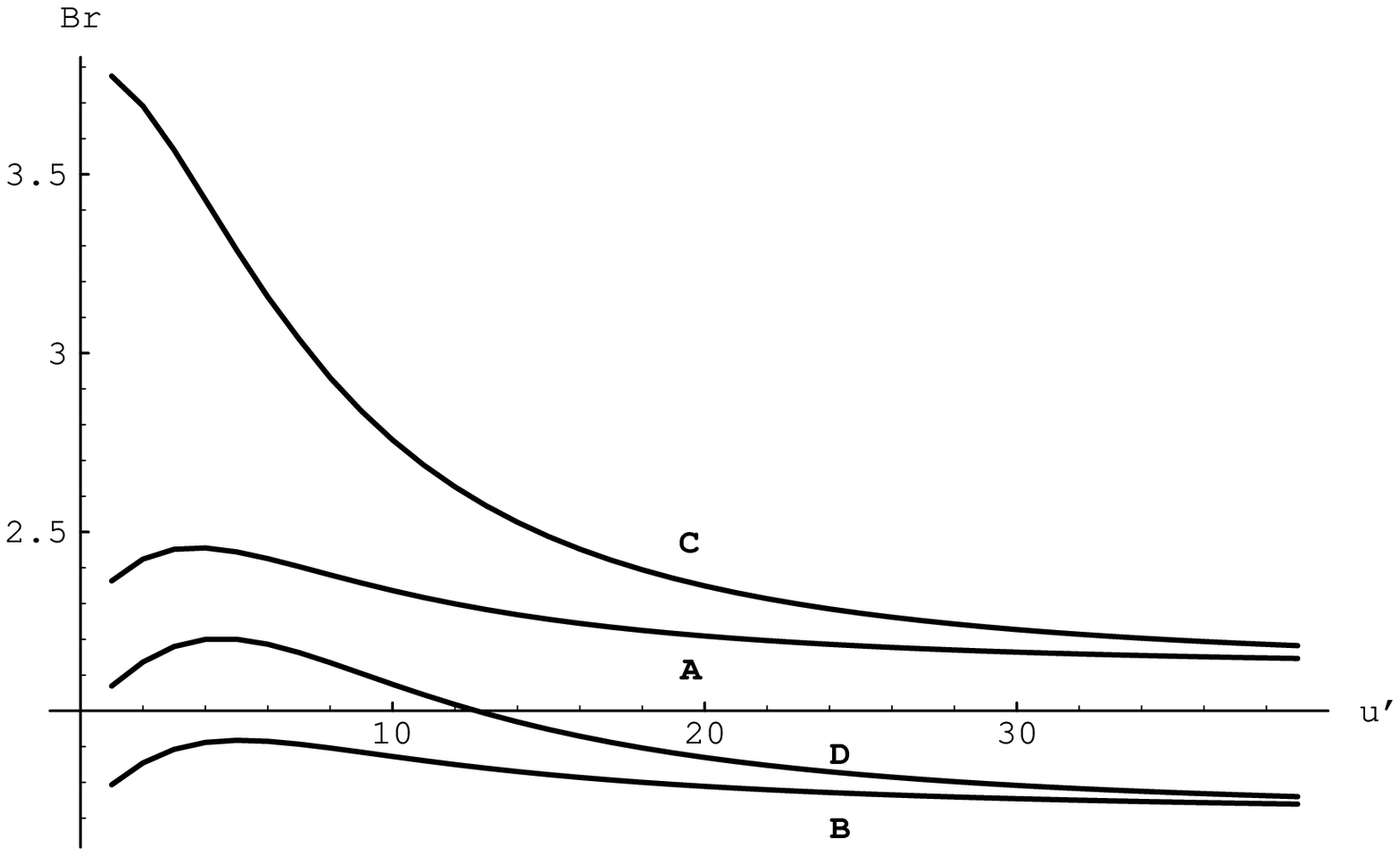}

{\Large\bf Fig. 2}
\end{center}
\end{figure}
\newpage
\begin{figure}[htbp]
\begin{center}
\epsfxsize=\textwidth
\epsfbox{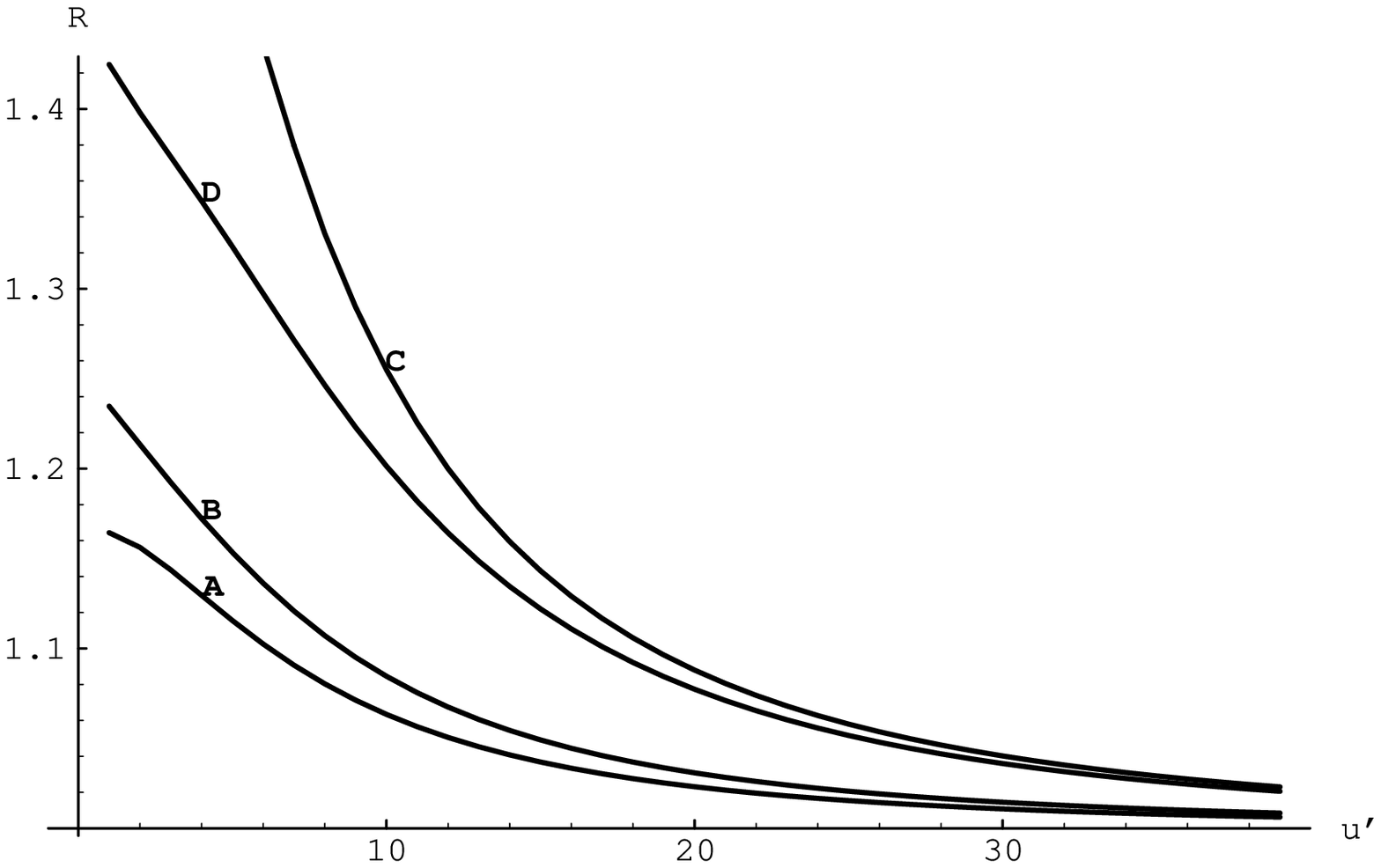}

{\Large\bf Fig. 3}
\end{center}
\end{figure}

\end{document}